\documentclass[apl,amsmath,amssymb,preprint]{revtex4-1}
\usepackage{graphicx}
\usepackage{braket}
\newcommand{\euler}[1]{{\usefont{U}{eur}{m}{n}#1}}
\newcommand{\eulerA}{\hbox{\euler{A}}}
\newcommand{\eulerB}{\hbox{\euler{B}}}
\newcommand{\eulerC}{\hbox{\euler{C}}}
\usepackage{hyperref}
\usepackage[version=3]{mhchem} % Formula subscripts using \ce{}

\begin{document}

\title{Relativistic origin of slow electron-hole recombination
in hybrid halide perovskite solar cells} 

% Place the author information here.  Please hand-code the contact
% information and notecalls; do *not* use \footnote commands.  Let the
% author contact information appear immediately below the author names
% as shown.  We would also prefer that you don't change the type-size
% settings shown here.

\author{Pooya Azarhoosh}
\affiliation{Department of Physics, Kings College London, London WC2R 2LS, UK}
\author{Jarvist M. Frost}
\affiliation{Centre for Sustainable Chemical Technologies and Department of Chemistry, University of Bath, Claverton Down, Bath BA2 7AY, UK}
\author{Scott McKechnie}
\affiliation{Department of Physics, Kings College London, London WC2R 2LS, UK}
\author{Aron Walsh}
\affiliation{Centre for Sustainable Chemical Technologies and Department of Chemistry, University of Bath, Claverton Down, Bath BA2 7AY, UK}
\affiliation{Global E$^3$ Institute and Department of Materials Science and Engineering, Yonsei University, Seoul 120-749, Korea}
\author{Mark van Schilfgaarde}
\email[Electronic mail:]{mark.van\_schilfgaarde@kcl.ac.uk}
\affiliation{Department of Physics, Kings College London, London WC2R 2LS, UK}

\date{\today}

%%%%%%%%%%%%%%%%% END OF PREAMBLE %%%%%%%%%%%%%%%%


\begin{abstract}
The hybrid perovskite \ce{CH3NH3PbI3} (MAPI) exhibits long minority-carrier lifetimes and diffusion lengths. We show that slow recombination originates from a spin-split indirect-gap. Large internal electric fields act on spin-orbit-coupled band extrema, shifting band-edges to inequivalent wavevectors, making the fundamental gap indirect. From a description of photoluminescence within the quasiparticle self-consistent \textit{GW} approximation for MAPI, CdTe and GaAs, we predict carrier lifetime as a function of light intensity and temperature. At operating conditions we find radiative recombination in MAPI is reduced by a factor of more than 350 compared to direct gap behavior. The indirect gap is retained with dynamic disorder.
\end{abstract}

\maketitle

% In setting up this template for *Science* papers, we've used both
% the \section* command and the \paragraph* command for topical
% divisions.  Which you use will of course depend on the type of paper
% you're writing.  Review Articles tend to have displayed headings, for
% which \section* is more appropriate; Research Articles, when they have
% formal topical divisions at all, tend to signal them with bold text
% that runs into the paragraph, for which \paragraph* is the right
% choice.  Either way, use the asterisk (*) modifier, as shown, to

Metal-organic perovskite solar cells, CH$_{3}$NH$_{3}$PbI$_{3}$ (MAPI) in
particular, have attracted much recent attention because of their high power
conversion efficiency and potential low cost.
The material exhibits strong absorptivity characteristic of a direct-gap
semiconductor, with the slow radiative recombination characteristic of an
indirect-gap semiconductor.
The minority carrier diffusion length considerably exceeds the material
thickness required for complete solar capture.
As such, internal quantum efficiencies approach 100\% \cite{Lin15}.
Power conversion efficiencies as high as 21\% have been reported\cite{zhou2014}.
%
% REF: new 21% record - http://advances.sciencemag.org/content/2/1/e1501170
%
The constituent elements are abundant and efficient devices can
be made with solution processing methods which offer the potential for low-cost
and large scale production.
MAPI is thus perhaps the first competitive realization of a ``third  
generation'' solar cell\cite{MGreen1982solar}.

Minority carrier recombination lifetimes of tens of microseconds are reported in MAPI
by time-resolved photoluminescent (TRPL) spectroscopy and other methods
\cite{Dong27022015,Shi30012015,Stranks2014}.
Such long lifetimes are found in high-quality samples of crystalline silicon,
the archetype indirect-gap semiconductor.
A variety of trap-based models have been used to interpret the TRPL
data\cite{Yamada2014,Stranks2014}.
These models suggest that the long lifetimes are due to immobilisation of
charges in traps.
However, samples (both single- and poly-crystalline) with trap density
differences of the order 10$^5$ have lifetime variations of only 
an order of magnitude.
This suggests that lifetime is weakly correlated with measured trap density.
Further, longer lifetimes are observed in low trap-density single-crystal samples.

We show that the observed slow radiative recombination is an
\emph{intrinsic property} of MAPI due to the details of the
electronic band structure.
We formulate the recombination rate in the framework of the quasiparticle
self-consistent \emph{GW} (QS\emph{GW}) approximation.
A spin-split indirect-gap is formed.  Assuming charge carriers thermalize
rapidly within a band, we calculate recombination as a function of charge
density (illumination intensity) and temperature.  
With low doping density and solar illumination intensities, the material
exhibits an indirect-gap. 
We show that under operating solar cell conditions, radiative recombination rate is
suppressed by over 350 fold; moreover, it varies in an anomalous
manner,
increasing rapidly with temperature, and illumination intensity and doping.
This is in contra-distinction to semiconductors such as CdTe or GaAs, whose
recombination properties are also calculated.

\begin{figure}[h!]
\begin{center}
%#ifdef pdflatex 
\includegraphics[width=\textwidth]{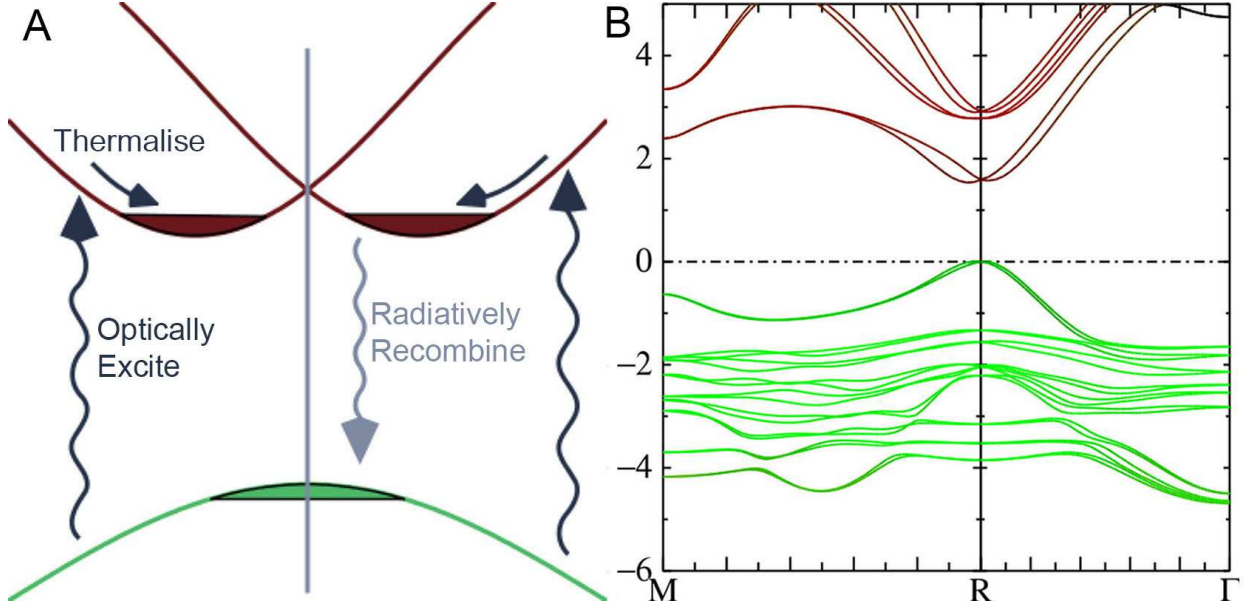}
%\includegraphics[height=3.75cm]{mapi-bands-near-r2.pdf}
%#else% 
%\includegraphics[height=3.75cm]{recombination.eps} \quad
%\includegraphics[height=3.75cm]{mapi-bands-near-r.eps} 
%#endif

\caption {%\small \raggedright
\textbf{Electronic band structure of CH$_3$NH$_3$PbI$_3$.}
({\textbf{A}}) schematic of absorption and recombination processes.
The conduction bands (red) are split around \textit{R}, as opposed to
conventional direct-gap semiconductors, where there is a single
minimum.  Photon absorption (dark grey arrows) generate electron-hole
pairs, which quickly thermalize to their respective band edges
(dark gray arrows) creating a quasi-equilibrium distribution of
carriers (red and green).  The excess electrons and hole
populations can recombine, creating a photon in the process
(light grey arrow).  Without phonon assist the transitions must
be vertical, as shown.  If the electron and hole populations
are small, the volume of overlapping $k$ space between the red
and green distributions becomes exponentially small in
1/$k_BT$, while in CdTe and GaAs the two populations occupy the same
phase space for any concentration.
\textbf{(B)} QS\emph{GW} band
structure for a section of the
$M$--$R$--$\Gamma$ lines.  There is a significant splitting of the
conduction band near $R$ ($E{\sim}1.75$\,eV), and a much smaller
splitting of the valence band, near ($E$=0).  }
\label{fig:keytran}
\end{center}
\end{figure}

For crystals without inversion symmetry, spin-orbit
coupling (SOC) splits spin-degenerate levels in non-magnetic systems.
In MAPI, significant
local electric fields, acting on the large SOC contribution from the heavy
lead atom, generate a significant shift in the spin-degenerate conduction
band minimum of Pb 6$p$ character.
This minimum splits into a pair of minima antipodal to the
original point\cite{brivio-2014}, causing the gap to become slightly indirect
(Fig. \ref{fig:keytran}).
We investigate the effect of dynamic disorder on this spin splitting.  Our
initial model is a single unit cell, with an infinite array of aligned
organic moieties.  In reality, the organic moiety rotates on a picosecond
timescale at room temperature
\cite{C4CP00569D,C4CC09944C,Leguy2015,Bakulin2015}.
We sample molecular dynamics realisations of a disordered supercell.
In spite of the disorder,
the spin-split indirect-gap is not only present but \textit{enhanced}.

The band splitting is a robust property of the material in working solar cells
that suppresses radiative recombination of minority charge carriers,
enhancing the photovoltaic action.
Two recent studies have noted the importance of the formation of an indirect band gap on the recombination rate of MAPI \cite{MottaRashba,RappeRecomb}. 
In Ref. \cite{MottaRashba}, an indirect gap was observed for certain molecular orientations.  
In Ref. \cite{RappeRecomb}, the indirect band gap is recognised as the result of Rashba splitting, leading to a mismatch in
both momentum and spin.  While their work emphasizes spin-mismatch, we
believe that the momentum-mismatch dominates device performance as the valence bands are
only slightly split, and geminate recombination is likely to be a minor process.

In this Letter, we report the recombination
rate with spin-orbit coupling included in the Hamiltonian, thus
directly treating both spin and momentum mismatch. To our knowledge, the only comparable calculation of radiative recombination was by Filipetti et al. where the rate was calculated from density-functional theory without the
spin-orbit coupling that is essential for both momentum and spin mismatch. 

Above 162 K, MAPI undergoes a continuous transition from a
tetragonal to a pseudocubic structure \cite{C4CC09944C}.
At room temperature the tetragonal distortion is small, $c/2a \approx 1.01$.
The organic moiety in MAPI has a relatively low barrier to rotation
\cite{FrostFerro}, and rotates at room temperature in a quasi-random fashion.
On average the structure can be considered to be cubic~\cite{C4CP00569D,C4CC09944C,Leguy2015}.
We initially construct a model with
a single formula unit of CH$_{3}$NH$_{3}$PbI$_{3}$.
We orient the organic moiety along [100], [110], and [111], where local
potential energy minima are found, which generate three structures
representative of the disordered system.
%From this configuration the atomic locations are energy minimized with the PBEsol density functional \cite{brivio-2013}.

% % % % % %Figure 1 goes here % % % % % % % 

In the absence of spin-orbit coupling the valence band maximum (VBM) and
conduction band minimum (CBM) $k_\mathrm{min}$ lie at the $\left< 111 \right>$ R point
\cite{brivio-2014,SpintronicAJF2014}.
The former consists mainly of I 5$p$ character, the latter mainly Pb
6$p$.
The degeneracy of both extrema are split and displaced by spin orbit
coupling ($\left<\xi(r)\right>\mathbf{L}\cdot\mathbf{S}$), which originates
near the atomic cores where $\xi(r)$ is large (Fig.~\ref{fig:keytran}).
$\left<\xi\right>$ scales approximately as $Z^2$.

With Pb being so heavy ($Z$=82), $\left<\xi\right>\mathbf{L}\cdot\mathbf{S}$
strongly affects the lead 6$p$ conduction
band, spin-splitting the minima into a pair of distinct states offset from
R (Fig.~\ref{fig:keytran}).
The iodine 5$p$ valence band is affected to a lesser extent ($Z$=53), the
spin-split displacement is small, resulting in a flattened region in reciprocal
space due to the overlapping minima.
The spin-orbit coupling narrows the band gap by $\sim$1\,eV. 
The spin splitting is linear in $k$, in contrast to typical semiconductors
where it varies as $k^3$.
% JMF: More explanation? Linear in k as in recip. space? What does this mean?
The band gap thus becomes slightly indirect.
The direct band gap $E_{0}$ remains at R, and is 75 meV larger than the
indirect gap.

The joint density-of-states (JDOS), relevant for absorption, differs from its 
direct-gap values in only a small energy range around $E_{0}$.
The flattened valence band contributes to a large density of states available
for optical transition at the direct-gap.
Thus \emph{absorption}  is only slightly affected by the
spin-splitting: MAPI absorbs solar radiation as though it were a direct-gap semiconductor.
However, \emph{radiative recombination} is dramatically suppressed. This is
extremely unusual for a solar cell material, where typically emission and
absorption are the direct reverse of one another.
Photoexcited electrons (holes) rapidly thermalize to a small region of $k$
space centered at the conduction (valence) band edges.
% The density of Fermi-Dirac distributions are shown diagrammatically in Fig.~\ref{fig:heatmap}.
The asymmetry in $k_\mathrm{min}^\mathrm{CBM}$ and
$k_\mathrm{min}^\mathrm{VBM}$ means that there is a low joint density-of-states
of the thermalized minority carriers.
Direct ($k$ conserving) recombination is thereby reduced.

Higher temperature leads to greater thermal broadening and a larger overlap in
the joint density-of-states, increasing radiative recombination.  
High density of photogenerated charges (or extrinsic charge carrier doping)
fill the small pockets at the CBM.  
Radiative recombination increases critically, reverting to direct-gap
semiconductor behaviour.
This property is \emph{intrinsic} to the material; no defect or trap states are
needed to explain why radiative recombination is suppressed at low photon flux, but increase dramtically at high photon flux.
% 
%Heavy illumination, extrinsic carrrier doping, or increasing temperature strongly enhances 
%the rate: as either electron or hole pocket fills up or spreads
%out the overlap between valence and conduction band pockets increases.  As the 
%carrier density becomes large the radiative decay dynamics 
%reverts to that of a conventional direct-gap semiconductor.
%
%Such phenomena have been observed experimentally as the 
There is some evidence for such behaviour: carrier lifetimes demonstrate
a sharp fall at moderate injection densities \cite{Yamada2014}.
% JMF: More qualitative description of literature here? Herz AFM 2015 paper? 

%\section{Ab initio description of Recombination}
%\label{sec:theory}

%\textbf{Ab initio description of recombination.}
%
Here we formulate an \emph{ab initio} theory for radiative
recombination within the QS\emph{GW} approximation\cite{lmsuitepackage}.
%It is noteworthy that we use QS\emph{GW} as opposed to more approximate schemes
%such as density functional theory within the Local Density Approximation (LDA).
 %or hybrid, density functional theory (DFT).                                                                                                                                                                      
QS\emph{GW} is parameter-free. 
In \emph{sp} semiconductors a wide range of electronic properties are uniformly
well described \cite{mark06qsgw}, including splitting from the Dresselhaus
terms \cite{PhysRevLett.96.086405}.
%This is essential for an electronic structure theory to be predictive in MAPI.
Few fully \emph{ab initio} formulations of radiative
recombination have been reported\cite{GLasherFStern}.  %semi-empirical 
Calculations using the van Roosbroeck-Shockley relation
%{\cred van Roosbroeck-Shockley relation; Pooya, please elaborateand we will put it in the supplemental material}                                                                                                  
with experimental\cite{BarugkinB} and theoretical\cite{AFlippettiB}
absorption coefficients  have been published.
We directly calculate the recombination rate without model parameters,
adapting the standard theory of dielectric response to non-equilibrium carrier
populations.

The recombination dynamics of carriers within the bulk of an intrinsic
semiconductor ($n{=}p$) can often be accurately described by a third-order rate
equation:
\begin{eqnarray}
    \frac{dn}{dt} = G - n\eulerA - n^2\eulerB - n^3\eulerC
    \label{eq:dynamics}
\end{eqnarray}
Here \emph{n} is the density of excited carriers and $G$ is a source term
describing constant photogeneration of carriers\cite{vanWalle}.
These parameters are often fit to experimental transient data across a large range of laser fluences and thus carrier
densities\cite{Milot2015}.
An implicit assumption is that these coefficients do not vary across the
carrier density regime experimentally accessed.
% %AND TEMEPRATURE
\euler{A} is related to the one-body non-radiative carrier recombination,
which proceeds through crystal defect levels as intermediate states.
We are concerned with intrinsic recombination and do not consider this
extrinsic process.
\euler{C} is the three-body Auger recombination coefficient.
This has been calculated in a \emph{GW} framework \cite{KotaniII} but it
becomes important under strongly non-equilibrium carrier populations and we omit
it here.
The two-body coefficient {\eulerB} describes radiative recombination of
free carriers and is intimately connected with both absorption and
emission.
Photo-generated carriers thermalize to the band edges on
a picosecond time scale~\cite{Price2015}, which is fast compared to
the radiative recombination time (ns to $\mu$s).
% JMF: For solar cells? 

We will assume that photoexcited carriers thermalize instantaneously
within a band to form quasi-equilibrium Fermi-Dirac distributions of electrons
and holes.
Thus the occupation probability of an excited electron \emph{c} is
given by $f_c{=}\left(\exp[(E_c{-}E_c^F)/k_BT]{-}1\right)^{-1}$, where
$E_c{-}E_c^F$ is the excitation energy relative to the electron quasi-Fermi
level.
The corresponding distribution $f_{v}$ for excited holes is the
same form, substituting $E_c{-}E_c^F{\to}E_v{-}E_v^F$.
In practice, we specify $T$, and electron and hole populations $n$ and $p$.
From the QS\emph{GW} band structure we can calculate the density of states and
so determine $E_c^F$ and $E_v^F$ from $n$ and $p$.

To establish that QS\emph{GW} can reliably predict the photoluminescent
process, we compute {\eulerB} for the benchmark materials GaAs and CdTe.
This both validates the method developed here and serves as points of
comparison to MAPI.

Under solar intensities a small population of electrons
(holes) is excited to the conduction (valence) band
(Fig.~\ref{fig:keytran}).
This density is relatively small for an operating solar cell
($\sim$ 10$^{17}$ cm$^{-3}$).
Only a small area of $k$ space near the band edges is utilized by photo-excited
charge carriers.
Numerically, this necessitates a fine $k$ mesh to adequately sample the near regions of the
band minima.
Some additional modest approximations are necessary to make the calculation tractable.
We neglect local fields and use the independent-particle (time-dependent
Hartree) approximation.
This approximation misses the Wannier excitons below the band
edge, but they are very shallow \cite{JackyEvenEb,Miyata2015}.
If the potential is local,
\begin{eqnarray}
    \epsilon(\omega) &=& \sum_{\mathbf{k}cv} \epsilon_{\mathbf{k}cv}(\omega) \\
    \hbox{Im}\,\epsilon_{\mathbf{k}cv}(\omega) &=&
    \frac{1}{\hbar} \left(\frac{2\pi e}{m\omega}\right)^2
    |P_{\mathbf{k}cv}|^2 f_c (1{-}f_v) \delta(\omega{-}\omega_{\mathbf{k}cv})\
    \label{eq:imeps}
\end{eqnarray}
$\epsilon_{\mathbf{k}cv}$ resolves $\epsilon$ into individual electron-hole
excitations between Bloch states \emph{v} and \emph{c};
$\hbar\omega_{\mathbf{k}cv}{=}E_c(\mathbf{k}){-}E_v(\mathbf{k})$ is the
excitation energy of the \emph{cv} pair.  The photons couple
\emph{c} to \emph{v}, which for direct transitions
simplifies to a matrix element of the momentum operator $P_{\mathbf{k}cv}$.
We omit phonon-assisted indirect transitions that do not conserve $k$
because they are weaker, higher-order processes.  
%Due to the only slightly indirect nature of the gap, the majority of radiative recombination will be by thermalization of the charge carriers to the band edge.
%
As written
Eq.~(\ref{eq:imeps}) is approximate because the QS\emph{GW}
potential is non-local.  However, for transitions close to $E_{0}$ the
effect of non-locality can be described by a scaling of $P_{\mathbf{k}cv}$ \cite{SoleandGirlanda}.
Further details of the theory are given in the
supplemental material \cite{supplement}.

By resolving $\hbox{Im}\,\epsilon$ into individual pair contributions,
the transition rate (Einstein coefficient) $A_{\mathbf{k}cv}$
between a \emph{cv} pair can be readily identified
\begin{equation}
    A_{\mathbf{k}cv}=\frac{n_re^2\omega_{\mathbf{k}cv}|P_{\mathbf{k}cv}|^2}{\pi\epsilon_0\hbar
    c^3m^2}
\end{equation} 
and the energy-resolved radiative recombination rate is
\begin{equation}
    R_{\mathbf{k}cv}(\omega) =  f_c (1-f_v)
    A_{\mathbf{k}cv}\delta[\omega-\omega_{\mathbf{k}cv}]
    \label{eq:transitionw}
\end{equation}
The total emission rate
\begin{equation}
    R^\mathrm{tot} = \int_0^\infty d\omega \sum_{\mathbf{k}cv}
    R_{\mathbf{k}cv}(\omega)
    \label{eq:transitiont}
\end{equation}
is conventionally expressed in terms of a carrier density independent
recombination lifetime $\tau$ or a
{\eulerB} coefficient:
\begin{equation}
    \tau^{-1}=\eulerB n=R^\mathrm{tot}/n
    . 
    \label{eq:lifetime}
\end{equation}

We neglect the scattering processes that eventually lead to thermalizstion of
electrons (holes) during photoluminescence. We adopt the standard approximation
in device modeling, and assume that recombination is slow compared to the
scattering of charge carriers within a band, so a quasi-equilibrium Fermi-Dirac
distrbution is maintained with a well-defined, band-dependent chemical
potential.

% % % % % Table 1 goes here % % % % % %

\begin{figure}[h]
\begin{center}             

\includegraphics[]{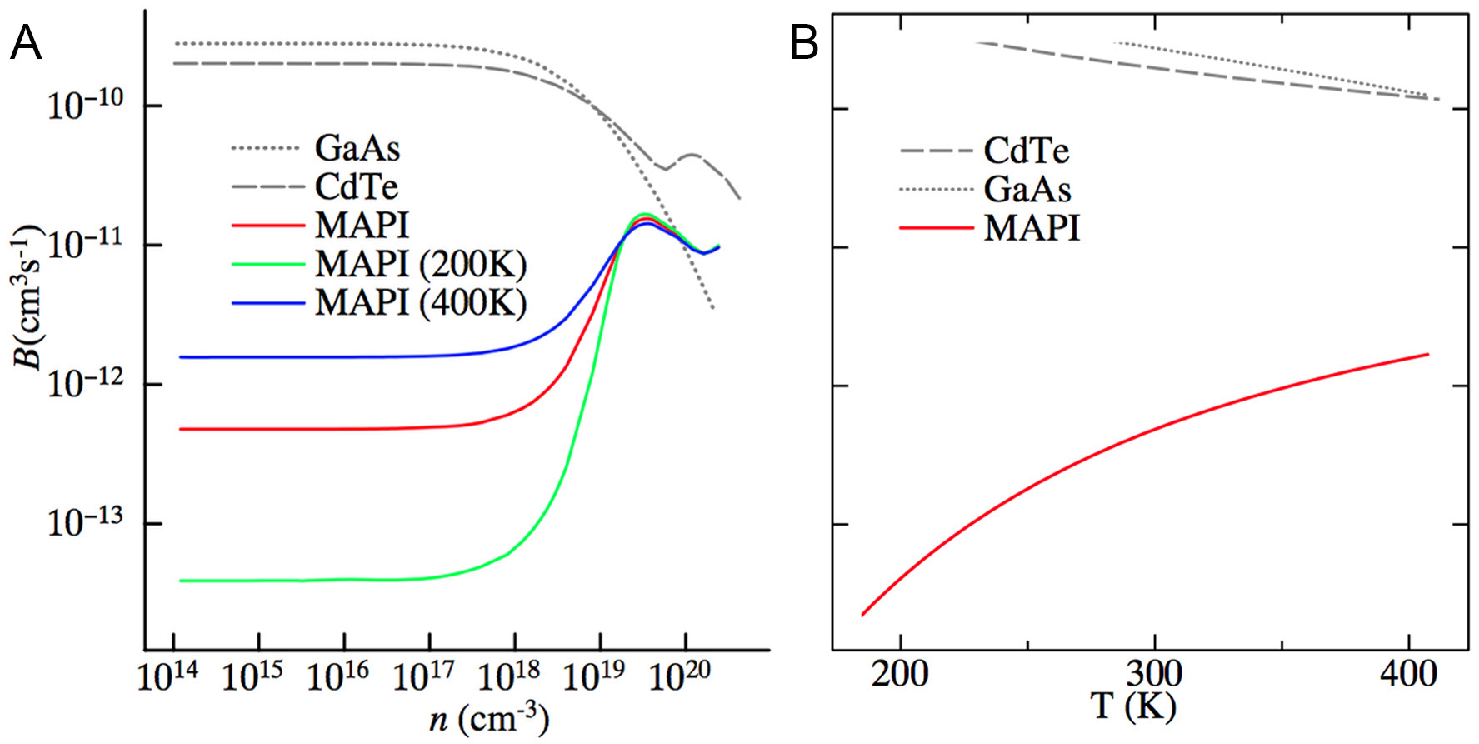}

\caption {\footnotesize %\raggedright
\textbf{Radiative recombination coefficient  for varying temperature and
excitation intensity.}
\textbf{(A)} Radiative recombination coefficient {\eulerB} in the intrinsic
case $n{=}p$.
Left: {\eulerB} for CdTe, GaAs and MAPI at room temperature, varying
photoexcitation density \emph{n}.  All have comparable direct
band gaps and effective masses.
At $n{>}10^{18}$cm$^{-3}$ the MAPI spin-split indirect-gap saturates
and the rate of radiative recombination  becomes similar in all three
materials.
Also shown are MAPI calculated at 200\,K and 400\,K.
The effect of the indirect gap becomes more pronounced as $T$ decreases.
\textbf{(B)} Temperature dependence of radiative recombination for
$n{=}10^{16}$\,cm$^{-3}$.}
\label{fig:Bcoeff}
\end{center}
\end{figure}

$\eulerB$, computed as a function of
carrier density for MAPI, CdTe and GaAs at several temperatures, is shown in Fig. \ref{fig:Bcoeff}.
Measured values at room temperature are displayed in Table \ref{Table:B-coeff}.
These vary widely, presumably owing to the competition of {\eulerB}
with other, non-radiative, recombination paths.
The presence of these pathways will depend strongly on how the material is
fabricated, and so the smallest measured $\eulerB$ value is probably the most
reliable, and the most directly comparable to our calculations. 
We calculate $\eulerB$ to be more than two orders of magnitude smaller
in MAPI than the direct-gap semiconductors CdTe and GaAs, but larger than the
fully indirect Si. 
MAPI resembles an indirect gap semiconductor for radiative carrier
recombination, and a direct gap semiconductor for absorption.

% % % % % %Figure 2 goes here % % % % % % %

The magnitude of {{\usefont{U}{eur}{m}{n} B}} increases as the \emph{k}-space
overlap between electron and hole distributions increases.  
This causes the rate of bimolecular recombination to depend on external
parameters in an unusual manner.  
For example, increasing temperature smears
$f_c$ and $f_{v}$ over a wider band of $k$ for fixed $n$, causing
{\eulerB} to increase with $T$.
The temperature dependence of {\eulerB} in GaAs and CdTe
(Fig.~\ref{fig:Bcoeff}) is weaker and of opposite sign.  
Further, bimolecular recombination will increase abruptly with photoexcited
carrier density once the electon pockets begin to fill up and overlap
($n {\sim} $10$^{18}$cm$^3$s$^{-1}$ in
Fig.~\ref{fig:Bcoeff}), in sharp contrast to CdTe and GaAs.  
Only at high carrier concentrations when a significant fraction of electrons and holes overlap 
% (Fig.~\ref{fig:heatmap}) 
does {\eulerB} in MAPI become comparable to {\eulerB} in CdTe, where it also
adopts the conventional behaviour and begins to decrease with increasing $n$.
%$n$ or $p$ can be increased either by external doping, or by increasing fluence.These data explain the necessity of reaching densities of $n{\sim}10^{18}$cm$^{-3}$ to observe stimulated emission in {\cred MAPI[REF]}.                                                                                                                                                                                                           
The strong carrier density, and therefore laser-fluence, dependence of
{\eulerB} suggests that global fits to time resolved data are not a reliable
method to infer {\eulerB} or lifetime from {\eulerB} (equation \ref{eq:lifetime}). 
%As {\eulerB} is a function of carrier density, you cannot directly
%invert {\eulerB} to derive a lifetime. 
The variation of {\eulerB} with carrier density will make high fluence
transients multi-exponential, and break the expected relationship between
light-emission and carrier-density sensitive experimental probes.

For photoexcitation densities $n {<} 10^{17}$cm$^3$s$^{-1}$
($n{=}p$), we find {\eulerB} $\approx$ 4$\times$10$^{-13}$ at room
temperature.
These are the charge carrier densities relevant for device operation in
sunlight. 
As {\eulerB} is fairly constant below this charge carrier density, we can use equation \ref{eq:lifetime} to derive a lifetime of order 10-100\,$\mu s$.
This lifetime is consistent with values
reported in literature for single crystal samples under 1 and 0.1 sun
intensity\cite{Dong27022015,Shi30012015,Stranks2014}.
% JMF: TODO ; this needs a little bit of discussion / understanding.
% Should introduce the AFM Herz paper + discuss.

There is limited temperature dependent TRPL data available.
Most data is for polycrystalline films, whereas our results would best be
compared to single crystal measurements.
Yet it has been observed that the carrier lifetime is highly temperature
sensitive at low fluence while being temperature insensitive at high fluence
\cite{Stranks2014}.
The increase in minority carrier lifetime with
decreasing sample temperature can be directly explained by a
contraction of the Fermi-Dirac distribution near the spin-split band minima.

%\textbf{Effect of dynamic disorder.}
%

So far we have considered MAPI in an idealized static structure.
In reality, the high temperature phase of MAPI is cubic on average only; the
dipolar molecules between cages continually rotate, the cages flex and tilt.  
Second-order Jahn-Teller deformations of the octahedra due to the Pb 6$s$ lone
pair directly distort the lead iodide bonds, generating local electric fields
near the atomic core region where spin-orbit coupling is high.
Recent molecular-dynamics simulations by Etienne et al. \cite{DynamicRashba} suggest that the
conduction band splitting persists in the presence of disorder.

If the material has true inversion symmetry, as MAPI is believed to in the
orthorhombic phase below 162 K, no directional electric fields can exist, and
so the spin-split indirect-gap should vanish. 
The rate of radiative recombination should increase, increasing the competition
with non-radiative recombination and thus increasing photoluminescent
efficiency. 

To assess the effect of dynamic disorder we performed \textit{ab-initio}
molecular dynamics at 300 K with a $2{\times}2{\times}2$ supercell.
%This is the smallest simulation cell that can enable the necessary octahedral
%tilting freedom (Brillouin zone boundary phonon modes).
Such supercells are able to accurately describe the phonon modes both at
the $\Gamma$ point and the Brillouin zone boundary; these $k$-points contain the important low-energy vibrational modes responsible for disorder.
We extract 100 realisations, each temporally separated by 0.25 ps, collected
after an initial equilibration period.
%After equilibration, we extracted 100 realisations, each temporally separated
%by 0.25\,ps.
%
Given the increased computational cost of this larger structure (96 versus 12
atoms), the electronic properties were calculated within the local density
approximation (LDA) including spin-orbit coupling.
While our intent is to show the qualitative effect of disorder on the
magnitude of the spin-split indirect-gap and the effective masses, the LDA
results for the ordered cells are in reasonable agreement
with QS\textit{GW} (Table \ref{Table:effectivemass}).

Between molecular dynamics snapshots, the spin-split band extrema and effective
masses fluctuate.
As can be seen from values averaged over the 100 instances 
(table \ref{Table:effectivemass}), disorder causes the light effective masses to increase.
This may explain the observed reduction in mobility as a function of
temperature, in the temperature regime where the organic moieties are
increasingly disordered.
Surprisingly, the band minima spin-split $k_\mathrm{min}$ \emph{also} increases with disorder.
This suggests that the cage deformation plays a more important role in
generating the directional electric field than the ordering of the moiety
dipoles.
This is indirectly supported by the success of the formamidinium based
perovskites, with smaller dipoles but larger anisotropy~\cite{FAPISnaith,FAPIJeon}.
                                                             
% % % % % % % % tabel 2 goes here % % % % % % % % %

%\section{Other mechanisms of reduced recombination}
%\textbf{Other mechanisms of reduced recombination.}
%

Although the model presented here shows good agreement with experiment,
other mechanisms that suppress bimolecular recombination may be operative.  
In particular, electric fields generated by local ferroelectric domains or
defect segregation could separate electrons and holes in \emph{real} space
\cite{FrostFerro,Lin-WangWangNanoLett}, in contrast to the present work,
where {\eulerB} is suppressed by separation in \emph{reciprocal} space.
Both mechanisms could contribute simultaneously to reduced recombination.  
We pointed to a limited body of evidence to support the latter; along these lines 
%We made testable predictions for its dependence on fluence, background doping,
%and temperature.  
it is noteworthy that high power conversion efficiencies have been reported
only for lead-based halide perovskites.  
This is consistent with the spin-split indirect-gap picture, as SOC is
weaker in lighter elements such as Sn.  In principle it is possible to
obtain experimental evidence for the direct/indirect-gap picture, with
optical probes of the (weakly emissive) indirect gap.  Sensitive
photoluminescence (PL) or electroluminescence (EL) would probe the
emission from this state, absorption can be probed by sensitive
external quantum efficiency (EQE) or photo-deflection spectroscopy
(PDS).  In our idealized structure we predict the splitting to be 75 meV,
while in the actual, disordered case it will vary with the disorder.
Thus we expect the splitting to decrease with temperature.

In summary we have provided a fully \emph{ab initio} relativistic calculation 
of the hybrid halide perovskite radiative recombination lifetime.
We considered only direct recombination, and with electron and hole populations
described by a quasi-Fermi level.
We have shown that the bimolecular recombination rate is strongly temperature
and carrier density sensitive, and that the long carrier lifetime and diffusion
length are a direct consequence of large relativistic spin-orbit coupling
combined with internal electric fields. 
We suggest that the relevant electric fields are mainly generated by dynamic
deformation of the inorganic octahedral cage. 
The spin-split indirect-gap is generated by the Pb lone-pair driven
distortions, and the Pb spin-orbit coupling of the conduction band.
For perovskite alloys, for example formamidinium/methylammonium/caesium and iodine/bromine,
the additional symmetry breaking due to occupational site disorder would be expected to further enhance
these effects, which is consistent with their strong photovoltaic action.
%This may explain why tin-halide and other perovskites are less efficient
%solar cells than their lead-based counterparts.

%The discovery and theoretical elucidation of this mechanism to reduce recombination can be applied to the discovery of new photovoltaic
%materials with similar performance to the halide perovskites.
%By having a heavy element with a high localisation of either the valence or
%conduction bands -- in a system which is capable of generating local electric
%fields (i.e. no centre of inversion) -- a spin-split indirect-gap can be
%generated.  
%Being able to tune the electronic band extrema in reciprocal space has potential in the design of materials with 
% extraordinary photovoltaic action.

\acknowledgments

The authors thank Jenny Nelson, Thomas Kirchartz and Keith Butler for useful discussions.
The research has been supported by the EPSRC (Grant Nos. EP/K016288/1,
EP/M009580/1 and  EP/M009602/1), the Royal Society, and the European Research
Council (No. 277757).  Computational resources were provided by the University
of Bath, and the EPSRC (Grant No. EP/L000202).

%Supplement

%Neglect of local fields: compare to dielectric functions paper.

%Approximation of velocity operator:
%When the effective potential is local, the velocity and momentum
%operators are proportional.  This does not apply to the nonlocal
%QS\emph{GW} potential; however if wave functions of some local potential
%are identical to the QS\emph{GW} potential, it is rigorously true that
%...  QS\emph{GW} and LDA eigenfunctions are not identical; this simple
%scaling factor yields some error as described in Ref. XX.  However, in
%the energy region around the gap (which will dominate recombination), the
%approximation is  a good one..

% Your references go at the end of the main text, and before the
% figures.  For this document we've used BibTeX, the .bib file
% scibib.bib, and the .bst file Science.bst.  The package scicite.sty
% was included to format the reference numbers according to *Science*
% style.

%\bibliography{plrefs2}
%\bibliographystyle{Science}

\begin{table}[h] \begin{center} \begin{tabular}{  l  l  l  p{6cm} } \hline
    \quad & Expt ($10^{-12}$cm$^{3}$-s$^{-1}$) & QS\emph{GW}  \\ \hline
    MAPI                 & see text            &  0.49  \\
    GaAs\cite{Hooft22,PJBishop1,FStern1}
          & 130--1300    & 267 \\
    CdTe\cite{CdTe1,CdTe2,CdTe3}  & 100--5100    & 195 \\
    %\hline                                                                                                                                                                                                        
    Si\cite{siliconB,siliconB2}  & 0.001--0.01 & \\  \hline
    \end{tabular}
        \caption{Experimental and QS\emph{GW} values for {\eulerB} at room                                                                                                                                         
    temperature, and $n{=}10^{17}$cm$^{-3}$.}
    \label{Table:B-coeff}
    \end{center}
    \end{table}

\begin{table}[h]
\begin{tabular}{ ccccccc}
  \hline
  \rule{0pt}{3ex} Structure & Theory & $k_{min}$ (\AA$^{-1}$) & $m_{1}$ & $m_{2}$ & $m_{3}$ & \\
  \hline\hline \vbox{\vskip 12pt}
  Pseudocubic $\left<100\right>$  & QS\textit{GW }& 0.043 & 0.95 & 0.13 &  0.11 &\\
  Pseudocubic $\left<110\right>$  &  & 0.055  & 0.13 & 0.11 &  0.11  &\\
  Pseudocubic $\left<111\right>$  &  & 0.043 & 0.36 & 0.12 &  0.12  &\\
  \hline\vbox{\vskip 12pt}
  Pseudocubic $\left<100\right>$  & LDA & 0.049  & 1.03 & 0.09 &  0.06 &\\
  Pseudocubic $\left<110\right>$  &  & 0.057  & 0.08 & 0.07 &  0.06  &\\
  Pseudocubic $\left<111\right>$  &  & 0.049  & 0.27 & 0.08 & 0.08 &\\
  \vbox{\vskip 6pt}
  Molecular Dynamics &  &0.105 & 1.02 & 0.60 & 0.42 &\\
  \hline
 \end{tabular}
 \caption{Calculated values of $k_\mathrm{min}$ and conduction band effective masses
  along the three principal axes. LDA results for the molecular dynamics
  $2{\times}2{\times}2$ supercell are the average from one hundred snapshots.}
\label{Table:effectivemass}
\end{table}

\clearpage

\newcommand{\beginsupplement}{%
        \setcounter{table}{0}
        \renewcommand{\thetable}{S\arabic{table}}%
        \setcounter{figure}{0}
        \renewcommand{\thefigure}{S\arabic{figure}}%
        \setcounter{equation}{0}
        \renewcommand{\theequation}{S\arabic{equation}}%    
     }

\beginsupplement

\section{Supplemental Material}

\subsection{Recombination Rate}\label{sec:LS}
Considering first order perturbation theory for the spontaneous emission from
a two state system, the Einstein \textit{A} coefficient is given by
%Perturbation theory for radiative transitions of a single atom is given by the expression for Einstein's A coefficients
%
\begin{equation}
A_{cv}(\omega_{cv})=\frac{n_re^2\omega_{cv}|P_{cv}|^2}{\pi\epsilon_0\hbar c^3m^2}\delta(E_c-E_v-\hbar\omega_{cv})\label{eq:EintA}
\end{equation}
Here $|P_{cv}|$ is the matrix element for this transition.
The states are labelled $c$ and $v$ in anticipation of these states referring
to conduction and valence bands of a periodic material.

Considering the continuum response of a medium, the imaginary component of the
dielectric function can be defined as 
\begin{equation}
\epsilon^{cv}_i=({2\pi e})^2\sum_{\vec{k}}|P_{cv}|^2\delta(E_c(\vec{k})-E_v(\vec{v})-\hbar\omega)
\label{eq:epsi}
\end{equation}
This can be directly related to the absorption through $\sigma(\omega)=\frac{4\pi\kappa(\omega)}{\lambda}=\frac{2\pi\epsilon(\omega)}{n_r\lambda}$.
Rearranging the momentum matrix element $P_{cv}$ and substituting into the imaginary component of the dielectric function, we can express the imaginary component of the dielectric function as a sum over Einstein coefficients\footnote{The Fermi functions$f_c(1-f_v)$ are added to equation \ref{eq:EintA} due to the transition from a two-state single atom picture to band structure.}
\begin{equation}
    \epsilon^{cv}_i(\omega_{cv}) 
    \frac{\omega_{cv}^{3}n_r}{4\pi\epsilon_0 \pi^2c^3\hbar}
    f_c(1-f_v)=
    \sum_k A_{cv} \: \delta[E_v(k)-E_c(k)-\hbar\omega_{cv}]\label{eq:vrs-1}
\end{equation}
Integrating over all transition energies, we get the total rate of transition per unit volume
for all direct transitions, such that:
\begin{equation}
R_{cv}^{Tot}=\frac{n_r}{4\pi^3c^3\hbar^4\epsilon_0}\int_0^\infty dE \epsilon^{cv}_i (E_{cv})E_{cv}^{3}=\int_{0}^{\infty}dE\sum_k A_{cv}\delta[E_v(k)-E_c(k)-\hbar\omega_{cv}]\label{eq:vrs0}
\end{equation}
In the limit of non density dependence, this quantity is related to the
radiative recombination coefficient $\eulerB$ and lifetime through
\begin{equation}
    \tau^{-1}=nB=\frac{R_{cv}^{Tot}}{n}
\end{equation}
In this work we directly evaluate $R_{cv}^{Tot}$ by Expression \ref{eq:vrs0} as we have
direct access to the microscopic optical matrix elements and imaginary part of
the dielectric function. 

This equation is equivalent to the van Roosbreck-Shockley formalism. 
This correspondence we will now show. 

We start by rewriting Equation \ref{eq:vrs-1} in terms of the absorption coefficient
\begin{eqnarray}
    \sigma(\omega)\frac{n_r\lambda}{2\pi} 
    \frac{\omega_{cv}^{3}n_r}{4\pi\epsilon_0 \pi^2c^3\hbar}
    f_c(1-f_v)&=&
    \sum_k A_{cv} \: \delta[E_v(k)-E_c(k)-\hbar\omega_{cv}]\\    
\sigma(\omega) 
\frac{[\hbar\omega_{cv}]^2n^2_r}{4\pi^3\epsilon_0c^2\hbar^3}
f_c(1-f_v)&=&
\sum_k A_{cv} \: \delta[E_v(k)-E_c(k)-\hbar\omega_{cv}]
\end{eqnarray}
Converting to cgs units we get the simplified expression
\begin{equation}
\sigma(\omega) 
\frac{[\hbar\omega_{cv}]^2n^2_r}{\pi^2c^2\hbar^3}
f_c(1-f_v)=
\sum_k A_{cv} \: \delta[E_v(k)-E_c(k)-\hbar\omega_{cv}]\label{eq:vrs1}
\end{equation}

If we integrate over all frequency and reciprocal space this equation is
equivalent to the van Roosbreck-Shockley (vRS) relation. 
The equivalence can be seen through the definition of the absorption coefficient. 
In this work we use an absorption coefficient for an athermal system with
completely empty conduction bands and completely full valence bands.
However one can define an absorption coefficient containing the thermal
excitation of charge carriers.  
This absorption coefficient $\alpha(\omega)$ is given by 
\begin{equation}
\alpha^{cv}_i(\omega_{cv}) =\frac{2\pi}{n\lambda} 
    \left( \frac{2\pi e}{m\omega_{cv}} \right) ^2 
    \sum_{\vec{k}}|P_{cv}|^2 \: (f_c-f_v)\delta(E_c(\vec{k})-E_v(\vec{k})-\hbar\omega)\label{eq:alpha}
\end{equation}
The Fermi functions in Equation \ref{eq:alpha} take into account stimulated emission and absorption. 
These two absorption coefficients are directly related through
$\sigma(\omega)=\frac{\alpha(\omega)}{(f_c-f_v)}$. 
We substitute this form into Equation \ref{eq:vrs1} to arrive at
\begin{eqnarray}
\alpha(\omega) 
\frac{[\hbar\omega_{cv}]^2n^2_r}{\pi^2c^2\hbar^3}
\frac{f_c(1-f_v)}{(f_c-f_v)}&=&
\sum_k A_{cv} \: \delta[E_v(k)-E_c(k)-\hbar\omega_{cv}]\\ 
\frac{[\hbar\omega_{cv}]^2n^2_r}{\pi^2c^2\hbar^3}\alpha(\omega)\frac{1}{\exp(\hbar\omega/k_BT)-1}&=&\sum_k A_{cv} \: \delta[E_v(k)-E_c(k)-\hbar\omega_{cv}]
\label{eq:vrs2}
\end{eqnarray} 
This is the conventional form of the van Roosbreck-Shockley relation, hence
showing the equivalence of Equation \ref{eq:vrs1} and \ref{eq:vrs2}. 

Our microscopic model for recombination formally agrees with the macroscopic
van Roosbreck-Shockley relation, in the limit where the photon bath within the
material is small, and so stimulated emission and absorption processes can be
neglected. 

\subsection{Effective optical matrix}\label{sec:DelSole}
The imaginary part of the dielectric function is shown in equation \ref{eq:epsi}. 
We can express the optical matrix element $P_{cv}$ in the momentum gauge as
\begin{equation}
    P_{cv}^{Loc} = 
    \frac{-i}{m \: \omega} 
    \bra{\psi_{c,\vec{k}}}\hat{e}\cdot{\vec{p}} \ket{\psi_{v,\vec{k}}}
    \label{eq:momentumgauge}
\end{equation}
This form has the advantage of taking $\lim_{q\rightarrow0}$ analytically. 
In the length gauge form
\begin{equation}
    P_{cv}^{Loc} = 
    \lim_{q\rightarrow0}
    \frac{1}{q}
    \bra{\psi_{c,\vec{k}+\vec{q}}}\exp(iq\hat{e}\cdot\vec{r}) \ket{\psi_{v,\vec{k}}}
\end{equation} 
the limit persists and must be taken numerically. 
Both these forms are equivalent. 
Formally, these are valid only for local Hamiltonians.  

This limitation can be overcome by scaling the matrix element by 
\begin{equation}
    P_{cv} = 
    P_{cv}^{Loc} S^* = 
    P_{cv}^{Loc} 
    \frac{E_c^{*}-E_v^{*}}{E_c^{0}-E_v^{0}}\label{eq:scale}
 \end{equation}
where $E_c^{*}$ and $E_v^{*}$ are the eigenvalues of the non-local Hamiltonian for the conduction and valence band respectively, while $E_c^{0}$ and $E_v^{0}$ are the local Hamiltonian counter parts.
 \\
 The derivation of expression \ref*{eq:scale} is outlined below: 
 \begin{eqnarray}
 \lim_{q\rightarrow0}\frac{1}{q}\bra{\psi^*{_c,\vec{k}+\vec{q}}}\exp(i\vec{q}\cdot\vec{r}) \ket{\psi^*_{v,\vec{k}}}&=&\lim_{q\rightarrow0}\frac{1}{q}\bra{\psi^*_{c}\ket{\psi^0_c}\bra{\psi^0_v}\ket{\psi^*_v}\bra{\psi^0_{c^,\vec{k}+\vec{q}}}\exp(i\vec{q}\cdot\vec{r}) \ket{\psi^0_{v,\vec{k}}}}\\
 &=&\frac{1}{m}\bra{\psi^*_c}\ket{\psi^0_c}\bra{\psi^*_v}\ket{\psi^0_v}\frac{\bra{\psi^0_c}\hat{e}\cdot\vec{p}\ket{\psi^0_v}}{E_c^{0}-E_v^{0}}\\
 &=&\frac{1}{m}\bra{\psi^*_c}\ket{\psi^0_c}\bra{\psi^*_v}\ket{\psi^0_v}\frac{\bra{\psi^0_c}\hat{e}\cdot\vec{p}\ket{\psi^0_v}}{E_c^{*}-E_v^{*}}\frac{E_c^{*}-E_v^{*}}{E_c^{0}-E_v^{0}}\\
  &=&\frac{1}{m}\bra{\psi^*_c}\ket{\psi^0_c}\bra{\psi^*_v}\ket{\psi^0_v}\frac{\bra{\psi^0_c}\hat{e}\cdot\vec{p}\ket{\psi^0_v}}{E_c^{*}-E_v^{*}}S^*=|P^*_{cv}|
 \end{eqnarray}
%
% TODO: JMF: i.e. is it a perturbation theory?
In the derivation above the asterixed symbols correspond to eigenfunctions $(\bra{\psi^*_c}\text{,} \bra{\psi^*_c})$ and eigenenergies $(E_c^{*}\text{,}E_v^{*})$ of a non-local Hamiltonian. 
This approximation is based on the assumption that the local and non-local
eigen functions are almost identical. 
Using this approximation the imaginary part of the dielectric function can be
approximated by
\begin{equation}
\epsilon^{cv}_i = 
\left( \frac{2\pi e}{m\omega} \right)^2
\sum_{\vec{k}}|P^*_{cv}|^2 \: 
\delta(E_c(\vec{k})-E_v(\vec{v})-\hbar\omega)
\end{equation}
This effective optical matrix has been developed by Sole and
Girlanda\cite{SoleandGirlanda}, and Levine and Allan\cite{LevineandAllan}
independently. 
We note that this approximation has no qualitative and little quantitative
effect on the central results of the paper. 
The long carrier lifetimes we predict are mainly due to the unusual joint
density of states calculated for these materials due to the spin-split
indirect-gap.
 %figure \ref{fig:DelSol} shows the validity of this approximation for small energy range above the band gap.
 \\
%\begin{figure}[h]
%\begin{center}
%\includegraphics[height=250pt]{DelSol.pdf}
%\caption{The imaginary part of the dielectric function for MAPI (C-N parallel to  $<100>$) using the effective optical matrix (Red) and the length gauge (Black).}\label{fig:DelSol}
%\end{center}
%\end{figure}
\\
%The effect of this approximation is shown in Fig\ref{fig:DelSol}, This shows that the effective matrix method is sufficient for the small energies above the band gap which is related to the recombination process described in the report.

\subsection{Computational procedure}
All calculations in this report were preformed using the \textsc{LMsuite}
codes\cite{lmsuite}, which implement an all-electron numeric solution of the
quasi-particle self-consistent GW approximation and the improved tetrahedron
method as formulated by Bl\"ochl {\it et al.}\cite{Tetra}.
The crystal structures used were energy minimised with the PBEsol\cite{PBEsol}
functional as implemented in the \textsc{VASP} codes\cite{VASP}. 
They are as previously reported\cite{brivio-2014}, and are available in
computer readable form~\cite{wmdgit}.  
The \textit{QS}GW band structures were calculated on
a $120\times120\times120$ k-mesh.
With this high resolution the quasi-Fermi levels are sampled with sufficient
accuracy. 
The partial densities of state were calculated for a given temperature and
carrier population with
\begin{equation}
n_c(E)=D(E)f(E,T)
\end{equation}
where $n_c(E)$ is the carrier population, $D(E)$ is the density of states and
$f(E,T)$ is the Fermi-Dirac function at a given energy and temperature. 
This assumes the full thermalisation of electrons and holes. 

The optical properties and emission spectrum are calculated using the methods
described in sections \ref{sec:LS} and \ref{sec:DelSole} and the original main
text of the article.  
For emission calculations only the lowest two conduction and highest two
valence bands are taken into account. 
An energy mesh of 3000 points was used between $E_g$ and $E_g+3.4$ eV for
emission spectrum calculations. 
This limit covered all emission, and was found to have sufficient mesh fineness
for the lowest carrier densities calculated. 

%\bibliography{plrefs2}


\end{document}